\documentclass[12pt,preprint]{aastex}

\begin{document} 
\newcommand{\met}{\hbox{E\kern-0.5em\lower-0.1ex\hbox{/}}_T}
\newcommand\simlt{\lower.5ex\hbox{$\; \buildrel < \over \sim \;$}}
\newcommand\simgt{\lower.5ex\hbox{$\; \buildrel > \over \sim \;$}}
    
\title{Variable TeV emission as a manifestation of jet formation in M87?}

\author{Amir Levinson$^1$ \& Frank Rieger$^2$}
\affil{$^1$ School of Physics and Astronomy, Tel Aviv University, 
Tel Aviv 69978, Israel\\
$^2$ Max-Planck-Institut f\"ur Kernphysik, PO Box 103980,
69029 Heidelberg, Germany}

\begin{abstract}
It is proposed that the variable TeV emission observed in M87 may be produced in a starved magnetospheric region, 
above which the outflow associated with the VLBA jet is established.   It is shown that annihilation of MeV
photons emitted by the radiative inefficient flow in the vicinity of the black hole, 
can lead to injection of seed charges on open magnetic field lines, with a density that depends sensitively on 
accretion rate, $n_\pm\propto\dot{m}^{4}$.  For an accretion rate that corresponds to the inferred jet 
power, and to a fit of the observed SED by an ADAF model, the density of injected pairs is found to be smaller
than the Goldreich-Julian density by a factor of a few.  
It is also shown that inverse Compton scattering of ambient photons by electrons (positrons) accelerating 
in the gap can lead to a large multiplicity, $\sim 10^3$, while still allowing photons at energies of up to a few 
TeV to freely escape the system.
The estimated gap width is not smaller than $0.01 r_s$ if the density of 
seed charges is below the Goldreich-Julian value.  The VHE power radiated by the gap can 
easily account for the luminosity of the TeV source detected by H.E.S.S.  
The strong dependence of injected pair density on accretion rate should render the gap emission highly intermittent. 
We also discuss briefly the application of this mechanism to Sgr A$^\star$.
\end{abstract}

\section{Introduction}

Combined VLBA and TeV observations of M87 reveal a rapidly varying TeV emission that appears to be 
associated with the m.a.s VLBA jet (Acciari et al. 2009).  The rapid flaring activity of the TeV source, with 
timescales $t=1t_{\rm day}$ day as low as 1-2 days, implies a source size of $d\sim 4r_s t_{\rm day}$ for a 
black hole mass $M_{BH}=4\times10^9$ solar masses.\footnote{While early results based on M87 gas 
kinematics suggested a black holes mass $M_{BH}=(3.2\pm0.9) \times 10^9 M_\sun$ (Macchettto et al. 
1997), recent modeling indicates $M_{BH}=(6.4\pm0.5) \times 10^9 M_\sun$ (Gebhardt \& Thomas 2009). 
For the following, a fiducial value of $4\times 10^9$ is adopted for M87.}

This, and the fact that the TeV emission appears to be correlated with the VLBA jet but not with emission 
from larger scales (and in particular HST-1), motivates the consideration that the observed TeV photons
originate from the black hole magnetosphere (Neronov \& Aharonian 2007; Rieger \& Aharonian 2008).   
A plausible magnetospheric process discussed in the literature is curvature and/or IC emission by particles, 
either hadrons or leptons, accelerating in a vacuum gap of a starved magnetosphere.  An alternative 
explanation for the observed fast TeV variability is emission from small regions located at larger radii, $r\sim 
100r_g$, as, e.g., in the misaligned minijets model of Giannios et al. (2010), or perhaps interaction of the 
jet with red giant stars closer to its base (Barkov et al. 2010).

The presence of the VLBA jet implies that a force-free (or ideal MHD) flow is established on scales $< 100 r_g$
(Walker et al. 2008, Acciari et al. 2009), so that the magnetosphere is anticipated to be screened in the sense 
that the invariant ${\bf E} \cdot {\bf B}$ nearly vanishes everywhere.   However, as in pulsar theory, there 
must be a plasma source that replenishes charges which escape the system (both, to infinity and across 
the horizon) along the open magnetic field lines in the polar region.   The nature of this plasma source is poorly 
understood at present.   

The injection of charges into the magnetosphere may be associated with the accretion process. Direct feeding 
seems unlikely, as charged particles would have to cross magnetic field lines on timescale shorter than the 
accretion time in order to reach the polar outflow.  Free neutrons that may be produced in a radiative inefficient 
accretion flow (RIAF) can cross field lines, however, they will decay over a distance $\sim 0.03r_s$ for a $10^9
M_\sun$ black hole, and even if existent at sufficient quantity, will not reach the inner regions.   On the other hand, 
MeV photons that are emitted by the hot gas near the horizon can annihilate in the polar region to produce charged 
leptons.   Below, it is shown that the density of the charges thereby injected depends sensitively on the accretion 
rate and the conditions in the RIAF.  Naive estimates suggest that in case of M87 this process cannot provide 
complete screening at accretion rates that corresponds to the inferred jet power, and to a fit of the observed SED 
by an ADAF model.  Those estimates are, however, highly uncertain, as explained below.

Another plausible plasma source is cascade formation in starved magnetospheric regions.  The size of the gap 
then depends on the conditions in the magnetosphere and the pair production opacity.   As shown elsewhere 
(Levinson 2000), vacuum breakdown by back-reaction is unlikely, as it requires magnetic field strength in excess 
of a few times $10^5$ G, higher than the equipartition value for Eddington accretion.  Pair production via absorption 
of TeV photons by the ambient radiation field is more likely.  However, the spectrum of the VHE photons observed 
by H.E.S.S. extends up to $\sim 10$ TeV (e.g., Aharonian et al. 2006), and the assumption that these photons originate 
from the magnetosphere (or even the VLBA jet) implies that the pair production opacity at these energies must not 
exceed unity.  This raises the question whether pair cascades in the magnetosphere can at all account for the 
multiplicity required to establish a force-free flow.

\begin{figure}
\centering
\includegraphics[width=140mm,height=100mm]{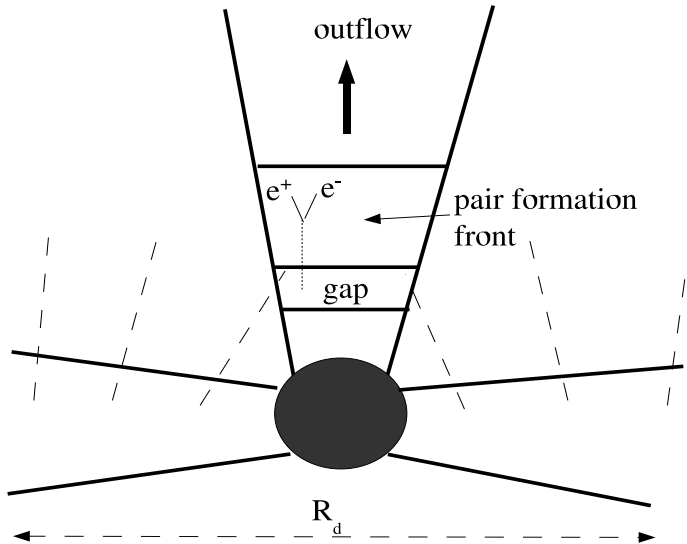}
\caption{\label{fig:a}Schematic representation of the magnetosphere structure: A vacuum gap of height $h<r_s$ accelerates particles (electrons or positrons) to high Lorentz factors.  The gap is exposed to soft radiation emitted from a source of size $R_d$.  Curvature emission and inverse Compton scattering of ambient radiation produce VHE photons with a spectrum extending up to $10^4$ TeV.   Photons having energies below a few TeV can escape freely to infinity.   Interactions of IC photons having energies well above 10 TeV with the ambient radiation initiate pair cascades just above the gap,  leading to a large multiplicity.  A force free outflow is established just above the pair formation front, and appears as the VLBA jet.   Intermittencies of the cascade process, induced by modest changes in accretion rate, 
give rise to the variability of the TeV emission observed by HESS, and the fluctuations of the resulted force-free outflow, as indicated
by the morphological changes of the VLBA jet. }
\end{figure}

In what follows it is shown that inverse Compton scattering (IC) of ambient photons by electrons (positrons) accelerating 
in the gap can lead to a large multiplicity, $\sim 10^3$, while still allowing photons at energies of up to a few 
TeV to freely escape the system.  The electromagnetic cascade is initiated by IC photons having much higher energies, 
up to $\sim 10^4$ TeV, for which the $\gamma\gamma$-optical depth is much larger.  The seed charges are provided by 
annihilation of MeV photons from the RIAF. It is found that the gap width is not smaller than $0.01 r_s$ if the density of 
seed charges is below the Goldreich-Julian (GJ) value.  The luminosity of the VHE photons produced in the gap can 
account for the TeV luminosity observed by H.E.S.S.  Any intermittencies of the cascade formation process would 
naturally lead to variability of both, the magnetospheric TeV emission and the resultant force-free flow, as observed.  
A schematic illustration of the model is presented in figure \ref{fig:a}. 

\section{The role of a radiative inefficient flow}
The strength of the magnetic field in a black hole magnetosphere is limited by the rate at which matter is accreted into 
the black hole.  At sufficiently low accretion rates the flow becomes radiative inefficient (RIAF) and the electron temperature 
in the inner region of the RIAF may exceed $m_ec^2$ (Narayan \& Yi 1995, hereafter NY95).  Bremsstrahlung cooling 
then gives rise to emission of soft gamma-ray photons that annihilate in the magnetosphere, leading to injection of charges 
on open magnetic field lines.   For sufficiently high annihilation rate the resultant charge density can exceed the GJ value 
everywhere, keeping the magnetosphere force-free.   At lower annihilation rates the magnetosphere will be starved.  In 
the latter case, the charge density produced by the annihilating photons from the RIAF defines the multiplicity required 
to establish a force-free outflow above the vacuum gap.  

In what follows we give a crude estimate of the ratio of the charge density produced by the RIAF and the GJ density, 
assuming that the accreting gas is in equipartition ($\beta=0.5$) with the magnetic field.

Henceforth, the accretion rate is measured in units of the Eddington rate, $\dot{m}=\dot{M}/\dot{M}_{Edd}$, where the 
Eddington accretion rate is defined as $\dot{M}_{Edd}=L_{Edd}/\eta_{ff}c^2=10^{27}M_9$ gr s$^{-1}$, with $\eta_{ff}=0.1$ 
adopted. The ion temperature of the accreted gas is close to virial, reaching $T_i\sim10^{12}$ K at $r=r_s$ .   The ion 
density is given by
\begin{equation}
n_i(r)=\frac{\dot{M}}{2\pi r^2 m_p v_r}=5\times10^{11}\dot{m}M_9^{-1}(r/r_s)^{-3/2}\quad{\rm cm^{-3}},\label{n_ion}
\end{equation}
adopting $v_r=0.1c(r/r_s)^{-1/2}$ for the RIAF radial velocity.   The Thompson optical depth is 
\begin{equation}\label{Thomson_depth}
\tau\simeq10^2 \dot{m}(r/r_s)^{-1/2}\,.
\end{equation}
Typically $\tau<<1$ for accretion rates $\dot{m}<<\dot{m}_{\rm crit}$, where $\dot{m}_{\rm crit}$ 
is the critical rate below which RIAF can exist.

We assume that the equipartition magnetic pressure is half the gas pressure, viz., $B^2/8\pi=0.5\rho_ic_s^2$, where 
$c_s\simeq c/\sqrt{3}\times (r/r_s)^{-1/2} $ is the sound speed.  This yields 
\begin{equation}
B\simeq 4\times10^4(\dot{m}/M_9)^{1/2}(r/r_s)^{-5/4}\qquad{\rm G}. \label{Beq}
\end{equation}
The numerical values in the expressions for $n_i$ and $B$ are in rough agreement with the results of  NY95 for a viscosity 
parameter $\alpha=0.3$,  adiabatic index of $4/3$, and radiative efficiency $1-f<<1$, where $f$ is the advection parameter, 
as defined in NY95.

At radii  $r<10^3 r_s$ the electron-ion coupling becomes weak by virtue of rapid cooling.   In this region electron cooling 
is dominated by synchrotron emission from thermal electrons.   For accretion rates near the critical rate $\dot{m}_{\rm crit}$ 
the electron temperature saturates at  $T_e\sim$ a few times $10^9$ K.  At such temperatures only photons emitted by 
electrons at the tail of the thermal distribution have energies above the pair production threshold.  However, for much 
lower accretion rates, $\dot{m}<<\dot{m}_{\rm crit}$, the electron temperature  approaches $10^{10}$ K, and thermal 
photons can annihilate.  

To estimate the annihilation rate at highly sub-critical accretion rates, we adopt the cooling functions for electron-ion and 
electron-electron Bremsstrahlung, $q_{ei}$ and $q_{ee}$, from NY95.  The total cooling rate per unit volume is then given 
to a good approximation by $q_{ff}=q_{ee}+q_{ei}\simeq7.5\times10^{-22}n_e^2\theta_e$ erg s$^{-1}$ cm$^{-3}$ at electron 
temperatures $\theta_e=k T_e/m_ec^2\simgt1$.  The numerical value corresponds to the choice $\theta_e=1$ in the 
logarithmic term in Eqs. (3.6) and (3.8) of NY95.  Since the RIAF is optically thin, pair production does not affect the leptonic 
content of the gas, and one can safely assume $n_e=n_i$.  By employing Equation (\ref{n_ion}) one arrives at
\begin{equation}
q_{ff}\simeq1.8\times10^{2}\theta_e\dot{m}^2M_9^{-2}(r/r_s)^{-3}\quad {\rm erg\ s^{-1}\ cm^{-3}}.
\end{equation}
The free-free luminosity emitted by the RIAF is $L_{ff}=\int{q_{ff}d^3r}\simeq 2\pi r^3 q_{ff} \ln(r/r_s)$, from which  
we readily obtain the number density of MeV photons in the magnetosphere:
\begin{equation}
n_\gamma=\frac{q_{ff}2\pi r^3 \ln(r/r_s)}{2\pi cr ^2\epsilon_{\gamma}}
                      \simeq \frac{0.2~q_{ff} r^3 }{c r_s ^2\epsilon_{\gamma}}
                      \simeq 1.4 \times 10^{11}\dot{m}^2M_9^{-1},
\end{equation}
where $\epsilon_\gamma=3\theta_e (m_e c^2)$.  The production rate of e$^\pm$ pairs inside the magnetosphere due to 
$\gamma\gamma$-annihilation is approximately $\sigma_{\gamma\gamma}n_\gamma^2c(4\pi/3)r_s^3$.  In steady state 
this rate is balanced by the escape rate, roughly $4\pi r_s^2 n_\pm c$.  Equating the two rates one has
\begin{equation}
n_\pm=\sigma_{\gamma\gamma}n_\gamma^2 r_s/3 \simeq 3 \times10^{11}\dot{m}^4M_9^{-1}
              \quad{\rm cm^{-3}}.\label{n_pm}
\end{equation}

The GJ density can be related to the accretion rate $\dot{m}$ through Eq. (\ref{Beq}) with $r=r_s$:
\begin{equation}
n_{GJ}=\frac{\Omega B}{2\pi ec}=5\times10^{-2}\dot{m}^{1/2}M_9^{-3/2}\quad{\rm cm^{-3}}.
\end{equation}
Thus we obtain
\begin{equation}
n_\pm/n_{GJ}\simeq 6 \times 10^{12}\dot{m}^{7/2}M_9^{1/2}.\label{n_pm-to-n_GJ}
\end{equation}
As seen, below a certain accretion rate, roughly $\dot{m}\simlt 2\times 10^{-4}$, injection of charges by this mechanism 
cannot provide complete screening of the magnetosphere.  The ratio $n_{GJ}/n_\pm$ then defines the 
multiplicity required to produce a force-free flow.   At higher accretion rates a vacuum gap may not exist.  However, as 
noted above, the electron temperature decreases with increasing $\dot{m}$, eventually dropping below $m_ec^2$.  
Detailed ADAF calculations in Kerr spacetime (e.g., Manmoto, 2000; Li et al. 2009) indicate a cutoff in the emitted 
spectrum at around $\theta_e\sim1$ even for $\dot{m}<<\dot{m}_{\rm crit}$.  For $\theta_e<1$ the number density of 
MeV photons is suppressed by a factor of roughly $6\exp(1/\theta_e)$, and $n_\pm\propto n_\gamma^2$ by a factor of 
$[6\exp(1/\theta_e)]^2$.  Thus, there is an uncertainty of two to three orders of magnitudes in our estimate of the pair 
density $n_\pm$.  This large uncertainty and the strong dependence of charge injection on accretion rate motivates 
detailed self-consistent calculations.   A complete, self-consistent treatment should also account for general relativistic 
effects, that are important close enough to the horizon.

The strong dependence on accretion rate suggests that a gap may form during periods of low accretion, so that emission 
from the gap may be intermittent.

\section{TeV emission and pair production from charges accelerating in a starved magnetosphere}

We consider a rapidly rotating black hole of mass $M=10^9M_9M_{\odot}$ embedded in an ambient radiation field.  
The radiation source is characterized by a luminosity $L_d=10^{41}L_{41}$ erg/s, a size $R_d={\cal R}r_s=
3\times10^{14}M_9{\cal R}$ cm, and SED (i.e., $\nu F_\nu$) peak energy $\epsilon_0=1\tilde{\epsilon}_0$ eV.
The energy density of this radiation field is $u_s\simeq3 L_{41}{\cal R}^{-2}M_9^{-2}$ erg cm$^{-3}$, and the 
corresponding number density of photons at the peak (measured over a logarithmic energy interval) is
\begin{equation}
n_s(\epsilon_0)\simeq1.8\times10^{12} L_{41}{\cal R}^{-2}M_9^{-2}\tilde{\epsilon}_0^{-1}\qquad {\rm cm^{-3}}.\label{n_s}
\end{equation}

\subsection{Curvature and inverse Compton emission}
The electric potential difference across a gap of height $h$ generated by a maximally  rotating black hole, can be 
expressed as 
\begin{equation}
\Delta V=1.7\times10^{21}B_4M_9(h/r_s)^2\quad {\rm Volts}.\label{BH-V}
\end{equation}
Charges accelerating in the gap will quickly reach a terminal Lorentz factor at which radiative losses balance energy gain,
viz., $q\Delta V= Pt=P h/c$ (Levinson 2000), where the net energy loss rate ,$P=P_{cur}+P_{IC}$, is the sum of  
curvature losses
\begin{equation}
P_{cur}=\frac{2}{3}\frac{q^2c\gamma^4}{\rho^2},\label{P_curv}
\end{equation}
here $\rho$ is the curvature radius of magnetic field lines, and inverse Compton (Thomson) losses
\begin{equation}
P_{IC}=\sigma_Tc\gamma^2u_s.\label{P_IC}
\end{equation}
 
Equating $\Delta V$ and $P_{curv}$ and using Eqs. (\ref{BH-V}) and (\ref{P_curv}), 
one finds that curvature radiation limits the Lorentz factor of emitting electrons (positrons) to
\begin{equation}
\gamma_{cur}=5\times10^{10}B_4^{1/4}M_9^{1/2}(h/r_s)^{1/4}(\rho/r_s)^{1/2}.\label{gam-curv}
\end{equation}
If the Thomson regime applies, inverse Compton scattering of electrons on the ambient photons limits the Lorentz factor 
to
\begin{equation}
\gamma_{IC}=\left(\frac{e\Delta V}{\sigma_Thu_s}\right)^{1/2}=2\times10^9
B_4^{1/2}M_9L^{-1/2}_{41}{\cal R}(h/r_s)^{1/2}.\label{gam-IC}
\end{equation}
Comparing Eqs. (\ref{gam-curv}) and (\ref{gam-IC}), it is seen that particle losses are dominated by inverse Compton 
scattering, viz., $\gamma_{IC}<\gamma_{cur}$, if 
\begin{equation}
L_{41}/{\cal R}^2>1.6\times10^{-3}B_4^{1/2}M_9(h/r_s)^{1/2}(\rho/r_s)^{-1}.  \label{comp}
\end{equation}

The spectrum of curvature emission peaks at an energy 
\begin{equation}
\epsilon_{cr,max}=\frac{3}{2}\frac{\hbar c\gamma^3_{\rm max}}{\rho}\le
10B_4^{3/4}M_9^{1/2}(h/r_s)^{3/4}(\rho/r_s)^{1/2}\qquad {\rm TeV},\label{e_max_curv}
\end{equation}
where $\gamma_{\rm max}={\rm min}(\gamma_{cur},\gamma_{IC})$.
The corresponding number of curvature photons emitted by a single particle is, to a good approximation, 
\begin{equation}
N_\gamma=P_{cur}h/(c\epsilon_{cr,max})\ge4\times10^{8}B_4^{1/4}M_9^{1/2}(h/r_s)^{5/4}
                       (\rho/r_s)^{-1/2}(1+f)^{-1},\label{n_cur}
\end{equation}
where $f=P_{IC}/P_{cur}$ denotes the ratio of IC and 
curvature loss rates.  Likewise, the maximum energy of IC photons is 
\begin{equation}\label{IC_max}
\epsilon_{IC,max}=m_ec^2\gamma_{\rm max}\le10^3B_4^{1/2}M_9L^{-1/2}_{41}{\cal R}(h/r_s)^{1/2} \qquad {\rm TeV}.
\end{equation}
The number of IC photons emitted by a single electron depends on the spectrum of the target radiation field. 
A crude estimate gives
\begin{equation}
N_\gamma=P_{IC}h/(c\epsilon_{IC,max})\ge10^{6.2}B_4^{1/2}L^{1/2}_{41}
                       {\cal R}^{-1}(h/r_s)^{3/2}(1+f^{-1})^{-1},\label{n_IC}
\end{equation}
but the actual number may be much larger, roughly by a factor of $m_ec^2/(\gamma_{max}h\nu_{s0})$ if the spectral 
energy distribution of the target radiation field peaks at a frequency $\nu_{s0}$ that satisfies  $h\nu_{s0}<<m_ec^2$. 
Inside the gap the field aligned electric field, $E_{||}$, is unscreened.  This implies that the charge density on magnetic 
field lines must not exceed the GJ value, $n_{GJ}=\Omega B\cos\theta/2\pi ec$.   The maximum gamma-ray power that 
can be produced by particles accelerating in the gap, regardless of the specific gamma ray production mechanism, is 
thus
\begin{equation}
L_{\gamma}=\int n_{GJ}c(e\Delta V) 2\pi r^2d\theta
                      \simeq3\times10^{47}\eta B_4^2M_9^2(h/r_s)^2 \qquad {\rm erg\ s^{-1}} ,\label{L_g}
\end{equation}
where $\eta$ is a geometrical factor. It is worth noting that the ratio $L_\gamma/L_{BZ}$, where $L_{BZ}$ is the maximum 
BZ power that can be extracted by a force-free flow,  scales as $(h/r_s)^2$.  If the pair density is well below the GJ density,
then the gamma-ray power in Eq. (\ref{L_g}) is reduced by a factor of $n_\pm/n_{GJ}$. 

\subsection{$\gamma\gamma$-opacity and pair multiplicity}

The soft photon energy that corresponds to the pair production threshold is $\epsilon_{\rm thr}=m_e^2c^4/\epsilon_\gamma$, 
which for peak curvature photons is 
\begin{equation}
\epsilon_{\rm thr}=2.5\times10^{-2}B_4^{-3/4}M_9^{-1/2}(h/r_s)^{-3/4}(\rho/r_s)^{-1/2}\qquad {\rm eV},
\end{equation}
and for IC photons is 
\begin{equation}
\epsilon_{\rm thr}=\frac{m_ec^2}{\gamma_{IC}} =2.5\times10^{-4} B_4^{-1/2}M_9^{-1}L^{1/2}_{41}{\cal R}^{-1}(h/r_s)^{-1/2}       
                                  \qquad {\rm eV}.
\end{equation}

Let us denote by $\zeta(\epsilon_s)$ the dimensionless energy distribution of soft photons, specifically the ratio of the number 
density at some energy $\epsilon_s$ and the number density of peak photons:
\begin{equation}
n_s(\epsilon_s)=\zeta(\epsilon_s) n_s(\epsilon_0).
\end{equation}
Then, invoking $\sigma_{\gamma\gamma}=0.2\sigma_T$ and using Eq. (\ref{n_s}), the pair-production optical depth to infinity 
at energy $\epsilon_s$ can be expressed as
\begin{equation}
\tau_{\gamma\gamma}=\sigma_{\gamma\gamma}n_s(\epsilon_s)R_d\simeq
70\zeta(\epsilon_s)L_{41}{\cal R}^{-1}M_9^{-1}\tilde{\epsilon}_0^{-1},\quad \epsilon_s\ge \epsilon_{\rm thr}.\label{taugg}
\end{equation}
The optical depth across the gap is smaller by a factor $h/R_d$, where we make the reasonable assumption that $h<R_d$.  
Note that the optical depth depends implicitly on the gap size $h/r_s$ through the value of $\zeta(\epsilon_{\rm thr})$ at the 
threshold energy.

Given $R_d$, $L_d$, $M_9$ and $B_4$ one can 
employ the above conditions in a self-consistent manner to solve for $h/r_s$ and $\gamma_{max}$.  
Let $dN_\gamma/d\epsilon_\gamma$ denote the number of photons per unit energy emitted by a single particle
in the energy interval $(\epsilon_\gamma,\epsilon_\gamma+d\epsilon_\gamma)$.  The gap multiplicity ${\cal M}$ can be defined
as 
\begin{equation}
{\cal M}=\int {\left(1-e^{[-(h/R_d)\tau_{\gamma\gamma}(\epsilon_\gamma)]}\right)\frac{dN_\gamma}{d\epsilon_\gamma}d\epsilon_\gamma},
\label{multiplicity1}
\end{equation}
and the gap size is then determined from the condition ${\cal M}=n_{GJ}/n_{\pm}$, where $n_\pm$ is the density of seed charges injected
into the open magnetosphere (see Eq. [\ref{n_pm-to-n_GJ}]).  If the latter condition cannot be satisfied it means that 
the gap must be fully restored, with $h/r_s\simeq 1$.   An outflow may still be established above the gap
if the size of the external radiation source is large enough, $R_d>>r_s\simeq h$.   In such a case the gamma ray photons emitted by the accelerating 
charges in the gap will interact with target photons at larger radii, $r>h$, ensuing pair cascades there.  The total multiplicity
can then be computed from Eq. (\ref{multiplicity1}) upon setting  $h/R_d=1$ in the exponent.  
 A force free flow will form above the starved magnetosphere provided ${\cal M}>n_\pm/n_{GJ}$.

\section{Application to M87 and Sgr A$^\star$}
\subsection{M87}
For this source we adopt a black hole mass of $M_9=4$.  
The assumption that the M87 jet is extracted magnetically by a BZ mechanism imposes a lower limit on the magnetic field 
in the vicinity of the horizon, and on the accretion rate. Various estimates (see, e.g., a summary in Li et al. 2009) yield a 
range of $L_j\simeq (10^{43} -  10^{44})$ erg/s for the power of the M87 jet.    
Equating $L_j$ with the BZ power that can be extracted from a Kerr black hole having a mass $M_9=4$ and specific angular 
momentum $a$, 
\begin{equation}
L_{BZ}=\frac{\epsilon}{64} a^2B^2r_s^2c\simeq3\times 10^{47}\epsilon a^2\dot{m} \qquad {\rm erg\ s^{-1}},\label{LBZ} 
\end{equation}
where $\epsilon\simlt1$ is an efficiency factor that depends on field topology and other details, implies an accretion rate $\dot{m}>10^{-4}\epsilon^{-1}a^{-2}$.  For a maximally rotating hole, $a=1$, the implied accretion rate is $\sim10^{-4}$, so
that the magnetic field strength close to $r_s$ becomes $B \sim 200$ G.  For an ADAF model with $L_b \sim 10~\dot{m}^2 
L_{\rm Edd}$ (NY95), this value for the accretion rate is consistent with limits on the bolometric luminosity measured in 
M87, $L_b \leq 10^{42}$ erg s$^{-1}$ (Owen et al. 2000). Note that the Bondi accretion rate in M87 inferred from Chandra 
measurements is $\dot{m}_B \sim 1.6 \times 10^{-3}$ (Di Matteo et al. 2003). The radial inflow velocity in a RIAF is about 
a factor $\alpha$ lower than in a Bondi flow, where $\alpha \sim 0.1$ is the viscosity parameter, so that one expects
$\dot{m} \sim \alpha \dot{m}_B$ (Narayan 2002), which would be consistent with the estimate above.

The condition (\ref{n_pm-to-n_GJ}) formally yields $n_\pm/n_{GJ}=0.12~(\dot{m}/10^{-4})^{7/2}$. However, as explained 
above this estimate is highly uncertain by virtue of the sensitive dependence of the injected pair density $n_\pm$ on the temperature of the accreting gas, on the geometry of the emitting region, and GR effects.  It seems likely, though, that a 
multiplicity of $10^2$ to $10^3$ should in any case be sufficient for complete screening.  

If sufficient charges cannot be provided by emission from the RIAF to ensure that  $n_\pm>n_{GJ}$, then a gap will exist, 
where the seed charges produced by annihilation of the RIAF photons accelerate.  These seed charges will emit TeV 
photons that can initiate pair cascades via interaction with the ambient radiation, as discussed above.  
However, observations constrain the $\gamma\gamma$-opacity at photon energies of a few TeV (e.g., Aharonian et al. 
2006), and so pair cascades can only be initiated by photons having energies well above 10 TeV.  In what follows, we 
employ some of the results derived in the preceding section to impose constraints on the maximum multiplicity and on 
the gap height.

Provided the RIAF SED is not significantly modified by the presence of an outflow, the emerging spectrum ranges 
from radio to X-ray energies. The radio to sub-mm regime is produced by synchrotron radiation from relativistic 
thermal electrons and dominated by emission from small radii. Compton up-scattering of the synchrotron photons 
fills in the sub-mm to hard X-ray regime, with most of the emission coming from the inner ($\simlt 10~r_s$) region
(e.g., Manmoto et al. 1997). For high $\dot{m}$, the Compton component is roughly a power law, while for low 
$\dot{m}$ distinct Compton peaks appear. This is related to the increase in electron temperature with decreasing 
accretion rates. At low $\dot{m}$, the X-ray spectrum is dominated by bremsstrahlung emission with a non-negligible 
contribution from large radii. 

The synchrotron emission is self-absorbed, resulting in a black-body spectrum up to frequency $\nu_{sa}$, where
it becomes optically-thin. In terms of ADAF parameters, one has
\begin{equation}
  \nu_{sa}(r)\simeq 9 \times 10^{10} M_9^{-1/2} (\dot{m}/10^{-3}\alpha)^{1/2}~T_{e,9}^2~(r_s/r)^{5/4}
  \quad \mathrm{Hz}\,,
\end{equation} where $T_{e,9}=T_e/10^9$K is the electron temperature (NY95; Mahadevan 1997). The luminosity 
then becomes (NY95; Mahadevan 1997) 
\begin{equation}\label{ADAF_peak_luminosity}
   \nu_{sa} L_{\nu_{sa}} \simeq 9 \times 10^{35} M_9^{6/5} \left(\frac{\dot{m}}{10^{-3}\alpha}\right)^{4/5}
                    \nu_{sa,11}^{7/5}T_{e,9}^{21/5} \quad \mathrm{erg/s}\,,
\end{equation} where $\nu_{11}=\nu/10^{11}$ Hz. As $T_e$ varies only slowly with $r$ in the inner region of an 
ADAF, the highest synchrotron emission frequency comes from the innermost part ($r\sim r_s$). For M87 with 
$M_9=4$, $\dot{m} \sim \alpha \dot{m}_B$ and $T_{e,9} \sim 5$ (cf. Mahadevan 1997), the synchrotron peak is 
located in the sub-mm band at
\begin{equation}
\nu_p = \nu_{sa}(r_s) \sim 10^{12} \;\mathrm{Hz}
\end{equation} and carries a characteristic power of  
\begin{equation}
 L_p = \nu_p L_{\nu_{p}} \sim 10^{41} \; \mathrm{erg/s}\,.
 \end{equation} This output would still comply to the upper limit imposed by Spitzer (MIPS)  observations 
 (several arcsec resolution), indicating that the (nuclear region plus galaxy) flux in M87 at $2 \times 10^{12}$ 
 Hz is $\sim (3-4) \times 10^{41}$ erg/s (Shi et al. 2007).

Compton up-scattering by a thermal distribution of relativistic electrons of small scattering depth $\tau \sim
10^{-2} (\dot{m}/10^{-4})~(r_s/r)^{1/2}$, Eq.~(\ref{Thomson_depth}), is characterized by a mean amplification 
factor per scattering of $A=1+4\theta+16\theta^2$. For ADAF temperature ranges of interest, typically $2 < A 
\simlt  10^2$. The frequency of the first (once-scattered) Compton peak is roughly $\nu_{c1} \sim A \nu_p$, 
and the power at this frequency is $L_{C1} \sim \eta_g L_p \tau A$ where $\eta_g\sim 0.5$ is a geometrical 
factor.  

Pair-production implies that TeV photons of energy $\epsilon$ produced in the magnetosphere will interact 
most efficiently with ambient soft photons of energy $\epsilon_s \sim (1~\mathrm{TeV}/\epsilon)$ eV. For 
moderate $A \sim 10$, the target soft photon field is dominated by the fraction of sub-mm photons scattered 
twice (2nd Compton peak) with relative peak flux $L_{C2}/L_p \sim A^2 \tau^2 \sim 10^{-2}$. Eq.~(\ref{taugg}) 
then suggests that $\tau_{\gamma\gamma}(\epsilon) <1$ up to VHE photon energies of several TeV. 

If, on the other hand, $A \gg 10$, then the energy of the once-Compton-scattered synchrotron photons can 
approach the infrared regime, $\nu_c \sim 10^{12} A$ Hz, yielding a non-negligible background for $\gamma
\gamma$-interaction with the TeV (curvature or IC) photons. Taking $L_{41}(\epsilon_s) \sim L_{C1}\sim 0.3$, 
Eq.~(\ref{taugg}) with ${\cal R} \sim 1$ formally implies that the pair-production optical depth for photons of 
$\epsilon \sim 1$ TeV is $\tau_{\gamma\gamma}(\epsilon) \sim 5$. However, this estimate is again somewhat 
uncertain by virtue of its sensitive dependence on accretion rate: 

The electron temperature in the inner region of an ADAF is only weakly dependent on accretion rate and at 
low $\dot{m}$ roughly follows $T_e \propto \dot{m}^{-q}$ with $0<q \simlt 0.2$ (Manmoto et al. 1997; 
Mahadevan 1997). The flux in the synchrotron peak, on the other hand, is highly dependent on the electron 
temperature and approximately varies as $L_p \propto \dot{m}^{3/2+3/4} T_e^7$ (Mahadevan 1997). For 
high enough temperatures $A \propto T_e^2$. This suggests that the Compton power, and thus the optical
depth may roughly vary with accretion rate as 
\begin{equation}
\tau_{\gamma\gamma}({\rm TeV}) \propto L_{C1} \propto \tau L_p A \propto \dot{m}^{2}\,.
\end{equation} An even more sensitive dependence $\tau_{\gamma\gamma}(\epsilon)\propto \dot{m}^{2.5-4}$ 
has been estimated based on detailed RIAF modelling of M87 (Li et al. 2009). Hence, only moderate changes 
in the accretion are required to make the source transparent to pair absorption up to photon energies of a few 
TeV as expected from IACT observations (e.g., Aharonian et al. 2006; Acciari et al. 2009). On the other hand, 
for energies well above 10 TeV severe absorption seems unavoidable.

The RIAF photons also provide a suitable target field for inverse Compton interactions with the energetic 
charged particles accelerated in the magnetosphere. As the energy loss rate for a single electron in the 
extreme Klein-Nishina limit is only weakly (logarithmically) dependent on the electron Lorentz factor, 
IC interactions (Thomson regime) with the radio-sub-mm disk photons (with the peak photons coming 
from $r \sim r_s$) becomes most constraining. IC interaction with photons of energy $\epsilon_p=h \nu_p$ 
occurs in the Thomson regime for electron Lorentz factors up to $\gamma_e \sim 10^8$. If the gap size is 
not too small ($h/r_s \simgt 10^{-2}$), electrons can get accelerated beyond this point ($P_{IC}(\gamma_e)~h/c 
< e \Delta V$). Taking $B_4\simeq 0.02$ and ${\cal R}\simeq 1$ for M87, IC scattering (\ref{gam-IC}) would 
limit achievable electron Lorentz factors to $ \gamma_{\rm IC} \simeq 10^9 L_{\rm t,41}^{-1/2} (h/r_s)^{1/2}$, 
where $L_{\rm t,41}=L_t/10^{41}$erg s$^{-1}$ is the luminosity of the target field for which Compton 
scattering occurs in the Thomson regime ($h \nu_t < m_e c^2/\gamma$). If one takes as a first approximation
\begin{equation}\label{target_field}
L_t (\nu_t) \simeq L_p \left(\frac{\nu_t}{\nu_p}\right)^{7/5}\,, 
\end{equation} as inferred from Eq.~(\ref{ADAF_peak_luminosity}), IC interactions would limit electrons 
Lorentz factors to 
\begin{equation}\label{gamma_IC}
     \gamma_{\rm IC} \simeq 4 \times 10^{11} L_{p,41}^{-5/3}(h/r_s)^{5/3}\,.
\end{equation} Note, however, that as $\nu L_{\nu} \propto \nu^{7/5}$, with $\nu \propto r^{-5/4}$, the 
dominant contribution to $L_t$ comes from larger scales, ${\cal R}_t\equiv{\cal R}(\nu_t) >1$, making the constraint 
even weaker, roughly by a factor of ${\cal R}_t^{10/3}$. 
Hence, if the gap is not too small, the energy loss of an electron will be dominated by curvature emission 
(\ref{gam-curv}), limiting achievable Lorentz factors to
\begin{equation}\label{gamma_c}
  \gamma_{c} \simeq 3.8 \times 10^{10} \left(\frac{h}{r_s}\right)^{1/4}\,,
\end{equation} where a curvature radius $\rho=r_s$ has been invoked for the magnetic field lines.   

Equation~(\ref{e_max_curv}) then indicates that the peak energy of the curvature spectrum is limited to $\sim50
\dot{m}^{3/8}$ TeV, which is below 1 TeV for $\dot{m}\simlt10^{-4}$. However, the maximum energy of IC 
photons up-scattered by the accelerated electrons exceeds $\gamma_c m_ec^2 \simeq10^{4}$ TeV. 
Given the expected spectral index of the radio disk spectrum in M87, cf. Eq.~(\ref{ADAF_peak_luminosity}), 
the average IC power emitted by a single electron in the Thomson regime; that is, due to scattering of photons
having $\nu<\nu_t=m_ec^2/\gamma_c$, is
\begin{equation}
P_{IC}(\nu_t)\simeq\frac{\gamma_c^2\sigma_TL_p}{2\pi R_d^2(\nu_t)}\left(\frac{m_ec^2}{\gamma_ch\nu_p}\right)^{7/5}
\simeq 3\times10^3L_{p,41}{\cal R}_t^{-2}(h/r_s)^{3/20}\quad {\rm erg/s}.
\end{equation}
These photons originate from a radius ${\cal R}={\cal R}_t>>1$.
The IC power emitted in the KN regime (for target photons at $\nu>\nu_t$) is reduced by a factor of $\ln(\nu_p/\nu_t)(\nu_t/\nu_p)^2$:
\begin{equation}
P_{IC}(\nu_p)\simeq\frac{\gamma_c^2\sigma_TL_p}{2\pi R_d^2(\nu_p)}\ln(\nu_p/\nu_t)(\nu_t/\nu_p)^2
\simeq 6\times10^2L_{p,41}{\cal R}_p^{-2}\quad {\rm erg/s},
\end{equation}
and depends only weakly (logarithmically) on the gap height.  These photons are emitted from ${\cal R}={\cal R}_p\sim1$.
The total power is $P_{IC}(\gamma_c)=P_{IC}(\nu_t)+P_{IC}(\nu_p)$.
The fraction of IC power to curvature power near the maximum energy, $f_m \equiv P_{IC}(\gamma_c)/
P_{cur}(\gamma_c)\simeq0.06L_{p,41}{\cal R}_p^{-2}(h/r_s)^{-1}[1+5({\cal R}_p/{\cal R}_t)^{2}(h/r_s)^{3/20}]$, is likely to 
exceed a few percent. 

The optical depth for the up-scattered photons at the highest energy $\epsilon_{IC}\simeq\gamma_cm_ec^2\simeq10^4$ TeV, cf. 
Eq.~(\ref{taugg}) with $\zeta(\epsilon_s)=(m_ec^2/\gamma_ch\nu_p)^{2/5}$, 
is likely to exceed $\tau_{\gamma\gamma}(\epsilon_{IC}) \sim 10^4$. Hence, these photons 
will be severely attenuated before escaping the source, producing pairs. If $(h/r_s)$ is not too small, this will 
happen inside the gap. The arising multiplicity is of order 

\begin{equation}
  N_{\gamma}(IC) = \frac{P_{IC} h/c}{\gamma_c m_e c^2}
                        \simeq 10^{2.9} \frac{L_{p,41}}{{\cal R}_p^{2}}
                        \left(\frac{h}{r_s}\right)^{3/4}[1+5({\cal R}_p/{\cal R}_t)^{2}(h/r_s)^{3/20}] \propto \dot{m}^{7/4-7q}\,,            
\end{equation}	
using $L_p \propto \dot{m}^{9/4-7q}$, $\gamma_c\propto  \dot{m}^{1/8}$. Typically, ${\cal R}_t>>{\cal R}_p$, 
so that the second term in the square parentheses can be neglected. 
The condition that the photon number $N_\gamma(IC)$ should exceed the multiplicity required to screen out 
the magnetosphere, ${\cal M}=n_{GJ}/n_\pm$,  where $n_\pm$ hereby refers to the seed charges, imposes 
a constraint on the gap height:
\begin{equation}
h/r_s >\left(\frac{{\cal M}}{10^{3}}\right)^{4/3} L_{p,41}^{-4/3} {\cal R}_p^{8/3}.\label{gap-h-min}
\end{equation} 
Note that the optical depth across the gap is $\sim \tau_{\gamma\gamma}(\epsilon_{IC}) h/R_d=\tau_{\gamma\gamma}(\epsilon_{IC})
(h/r_s){\cal R}^{-1}$ and therefore larger than unity if $h/r_s \simgt 10^{-4}{\cal R}$. 

For the above choice of parameters, equation (\ref{L_g}) yields a gamma-ray  luminosity of $L_\gamma\sim10^{45}
\eta(h/r_s)^2$ erg/s, which for $\eta=0.1$, $(h/r_s)^2\simgt10^{-3}$ is in excess of the observed TeV luminosity.  
The sensitive dependence of seed charges from the RIAF emission on the accretion rate strongly suggests that 
moderate changes in accretion rate will induce nonlinear variations  of the gap height, and conceivably its 
occasional disappearance.  This will result in intermittencies of the gap emission and the resultant outflow.

While the curvature emission is anticipated to peak at sub-TeV energy (although any spread in the curvature radius of
magnetic field lines will smear out the peak), the emitted spectrum at higher energies 
should depend on the details of the pair cascade process.    The cascading gamma rays will be emitted from their local 
photosphere, at an energy that varies with radius.  As demonstrated in Blandfrod \& Levinson (1995), inhomogeneities can 
lead to a very broad spectrum even in case of mono-energetic injection of electrons.  Thus, detailed calculations of the cascade 
process are required to determine the shape of spectrum emitted from the vicinity of the gap. 

\subsection{Sgr A$^\star$}
Estimates of the accretion rate that accommodate the extraordinary low bolometric luminosity of Sgr A$^\star$, $L_b<
10^{-8}L_{Edd}$, are in the range $\dot{m}\sim (0.1-5)\times10^{-5}$, adopting a black hole mass $M_9=4\times 10^{-3}$
(e.g., Narayan 2002; Nayakshin 2005; Cuadra et al. 2006). VHE observations of the Sgr~A$^\star$ region indicate a
gamma-ray spectrum that can be described by a power-law extending beyond $10$ TeV above threshold of $165$ GeV,
with an associated VHE luminosity of a few $10^{35}$ erg/s (Aharonian et al. 2004). 
The source of this emission, whether associated with the putative BH or with a different object, e.g.,
the PWN G359.95-0.04, is unresolved yet.  From  Eq. (\ref{n_pm-to-n_GJ}) we 
obtain $n_\pm/n_{GJ}\simeq10^{-6}~(\dot{m}/10^{-5})^{7/2}$.  Equation (\ref{comp}) implies that energy losses are 
dominated by curvature emission as long as ${\cal R}>2 ~(\dot{m}/10^{-5})^{-1/8}L_{36}^{1/2}$, where $L_{36}=1$ is 
roughly the bolometric luminosity observed in Sgr A$^\star$, which is dominated by emission at frequencies around the 
SED peak, $\nu_p\sim10^{12}$ Hz.
The potential drop across the gap is limited to $\Delta V\simeq10^{18}(\dot{m}/10^{-5})^{1/2}$ V, and from Eq. (\ref{gam-curv})
we deduce that the Lorentz factor of accelerating electrons is $\gamma\sim 2 \times 10^9(\dot{m}/10^{-5})^{-1/8}$. 
Thus, the peak energy of curvature photons is $\epsilon_{cr}\sim200(\dot{m}/10^{-5})^{3/8}$ GeV, and the maximum
energy of IC scattered photons $\epsilon_{IC}$ is not expected to exceed $\gamma m_ec^2\simeq 10^3$ TeV.  The pair 
production optical depth at this energy is $\tau_{\gamma\gamma}(\epsilon_{IC})\sim10^2{\cal R}^{-1}$, so the IC scattered 
photons are likely to be converted inside the gap.  Equations (\ref{Beq}) and  (\ref{n_IC}) with $L_{41}=10^{-5}$ yield 
$N_\gamma\sim10^{3.5}(\dot{m}/10^{-5})^{1/4}{\cal R}^{-1}$, from which we infer that multiplicity may be high enough to 
allow the pair density to approach the GJ density if $\dot{m}>10^{-4}$ or so.  For lower accretion rates we can safely assume 
$h\simeq r_s$. The present estimates of $\dot{m}$ imply that a force-free flow cannot be established by this mechanism.

The gap will, nonetheless, emit gamma-rays with a total luminosity
\begin{equation}
L_\gamma\simeq10^{35}(\dot{m}/10^{-5})^{9/2}\eta \quad {\rm erg/s}.\label{SgrA-Lg}
\end{equation} 
This emission is anticipated to peak in the sub-TeV band, at $\sim 200$ GeV.  Some fraction may be emitted also at
much higher energies.

\section{Summary}
Annihilation of MeV photons that are emitted by a radiative inefficient flow in the vicinity of a supermassive 
black hole leads to injection of charges into the black hole magnetosphere.  The density of the injected charges depends 
sensitively on the accretion rate. At sufficiently large accretion rates the density may exceed the GJ density, giving rise to a 
complete screening of the field aligned electric field, and the formation of a force-free outflow.  At lower accretion 
rates complete screening cannot be accomplished and a gap forms.  The seed charges injected into the gap are quickly 
accelerated along magnetic field lines by the large potential drop generated by the rotating hole, until reaching a  Lorentz 
factor at which energy gain is balanced by curvature and inverse Compton losses. The curvature and IC spectra peak in 
the TeV band.  The interaction of the TeV photons thereby produced with the ambient radiation field may give rise to 
initiation of pair cascades, with a multiplicity that depends on the accretion rate.

The proposed scenario is applicable to variable TeV-emitting Galactic Nuclei that accrete at sufficiently low rates. Out of 
the nearby under-luminous TeV candidate sources (Sgr~A$^\star$, Cen~A, M87, NGC~1399), M87 and Sgr~A$^\star$ 
are particularly interesting:\footnote{NGC~1399 has not yet been detected at TeV energies, while Cen~A may not 
accrete at sufficiently low rates.}

In case of M87 we estimate (with a large uncertainty)  that annihilation of MeV photons from the RIAF cannot provide 
complete screening for accretion rates that correspond to the inferred jet power, and to a fit of the observed SED 
by an ADAF model.   However,  for a fully restored gap a large multiplicity, in excess of $10^3$, is expected 
owing to absorption of $>>10$ TeV photons that are produced through inverse Compton emission of the charges 
accelerating in the gap.  The pair production opacity decreases with decreasing gamma-ray energy, and becomes 
sufficiently small below $10$ TeV to allow photons at these energies to freely escape the system.  We propose that the
variable TeV emission detected by HESS was emitted from a magnetospheric gap located at the base 
of the VLBA jet.  We note that rapid variability is naturally accounted for by this mechanism, as even moderate changes 
in accretion rate will lead to nonlinear fluctuations of the gap potential and the resultant TeV emission, 
owing to the sensitive dependence of the density of injected charges on accretion rate.   The observed gamma-ray spectrum 
should depend on the spectrum of soft (scattered) photons and the pair cascade process, and can, in fact, be quite broad.    
Moreover, any spread in the curvature radius of magnetic field lines in the gap will lead to a spread in the maximum 
energy of accelerated pairs, which in turn broaden the spectrum further.   Detailed calculations are needed in order to 
determine the shape of the emitted gamma-ray spectrum. Additional TeV photons may be emitted by the jet on larger scales, 
but this component should be more steady.  It can be useful to trace and disentangle the spectrum of the variable component 
alone. 

Our analysis indicates that pair production alone would lead to a very high-sigma flow, which doesn't
seem to be supported by the observations.  Other processes, specifically magnetic field annihilation
of a striped wind configuration by a Rayleigh-Taylor instability (Lyubarsky, 2010), might give rise to additional
pair creation during the acceleration phase of the outflow.  This, however, would still imply
asymptotic Lorentz factor well in excess of that inferred from observations (albeit on much larger scales).  Mass loading of 
the outflow may be accomplished through entrainment of surrounding matter, e.g., due to a non-dipolar magnetic field
(McKinney \& Blandford, 2009), or some instability of the wall jet interface.  
The resulting structure may consist of a fast (high-sigma) flow, surrounded by a slower sheath (e.g., Levinson \& Eichler 1993). 
The very inner magnetosphere may still support a vacuum gap if entrainment occurs on scales larger than several 
$r_g$.  The details of such putative structures remain to be explored.

In case of Sgr A$^\star$ we find that for current estimates of the accretion rate the density of pairs in the magnetosphere is 
likely to be below the GJ value.   Gamma-ray emission from the gap is expected with a total luminosity that depends sensitively 
on accretion rate, Eq. (\ref{SgrA-Lg}), and an SED peak in the sub-TeV band. 

We thank John Kirk, Yuri Lyubarsky and Jon McKinney for enlightening discussions.  AL acknowledges support by 
an ISF grant for the Israeli Center for High Energy Astrophysics.
F.R. acknowledges support by a LEA Fellowship.

\break


\begin{thebibliography}{99} 
\bibitem{BL1} Acciari et al. 2009, Science 325, 444
\bibitem{BL1a} Aharonian, F. et al. 2004, A\&A 425, L13
\bibitem{BL2} Aharonian, F. et al. 2006, Nature 314, 1424
\bibitem{BL3} Barkov M.V., Aharonian, F.A., Bosch-Ramon, V. 2010, ApJ in press (arXiv:1005.5252)
\bibitem{BL4a} Blandford, R. D. \& Levinson, A. 1995, ApJ, 441, 79
\bibitem{BL4b} Cuadra, J. et al. 2006, MNRAS 366, 358
\bibitem{BL4} Di Matteo, T. et al. 2003, ApJ 582, 133
\bibitem{BL5} Gebhardt, K. \& Thomas, J. 2009, ApJ 700, 1690
\bibitem{BL6} Giannios, D. et al. 2010, MNRAS 402, 1649 
\bibitem{BL7} Levinson, A. 2000, PRL 85, 912
\bibitem{LE93} Levinson, A. \& Eichler, D. 1993, ApJ, 418, 386
\bibitem{BL8} Li, Y.-R. et al. 2009, ApJ 699, 513 
\bibitem{Lyu10} Lyubarsky, Y. 2010, ApJ 724, 234 
\bibitem{BL9} Mahadevan, R. 1997, ApJ 477, 585
\bibitem{BL10} Manmoto, T. et al. 1997, ApJ 489, 791
\bibitem{BL11} Manmoto, T. 2000, ApJ, 534, 734
\bibitem{MB09} McKinney, J. C. \& Blandford, R. D. 2009, MNRAS, 394, L126
\bibitem{BL12} Narayan, R. \& Yi, I. 1995, ApJ 452, 710 (NY95)
\bibitem{BL12b} Nayakshin, S. 2005, A\&A 429, L33
\bibitem{BL13} Neronov, A., Aharonian, F.A. 2007, ApJ 671, 85
\bibitem{BL14} Narayan, R. 2002, arXiv:astro-ph/0201260
\bibitem{BL15} Owen, F.N. et al. 2000, ApJ 543, 611
\bibitem{BL16} Rieger, F.M., Aharonian F.A. 2008, A\&A 479, L5
\bibitem{BL17} Shi, Y. et al. 2007, ApJ 655, 781
\bibitem{BL18} Walker, R.C. et al. 2008, JPhCS 131, 2053



\end{thebibliography}
\end{document}